# Cognitive MIMO Radio: A Competitive Optimality Design Based on Subspace Projections

Gesualdo Scutari[1], Daniel P. Palomar[2], and Sergio Barbarossa[1]

E-mail: {scutari, sergio}@infocom.uniroma1.it, palomar@ust.hk

[1] Dpt. INFOCOM, Univ. of Rome "La Sapienza", Via Eudossiana 18, 00184 Rome, Italy

[2] Dpt. of Electronic and Computer Eng., Hong Kong Univ. of Science and Technology, Hong Kong.

## 1 Introduction

Radio regulatory bodies are recently recognizing that rigid spectrum assignment granting exclusive use to licensed services is highly inefficient., due to the high variability of the traffic statistics across time, space, and frequency. Recent Federal Communications Commission (FCC) measurements show that, in fact, the spectrum usage is typically concentrated over certain portions of the spectrum, while a significant amount of the licensed bands (or idle slots in static time division multiple access systems with bursty traffic) remains unused or underutilized for ninety percent of time [1]. It is not surprising then that this inefficiency is motivating a flurry of research activities in engineering, economics and regulation communities in the effort of finding more efficient spectrum management policies.

As pointed out in many recent works [2, 3, 4, 5], the most appropriate approach to tackle the great spectrum variability as a function of time and space calls for *dynamic* access strategies that adapt to the electromagnetic environment. Cognitive Radio (CR) originated as a possible solution to this problem [6] obtained by endowing the radio nodes with "cognitive capabilities", e.g., the ability to sense the electromagnetic environment, make short term predictions, and react intelligently in order to optimize the usage of the available resources. Multiple paradigms associated with CR have been proposed [2, 3, 4, 5], depending on the policy to be followed with respect to the *licensed* users, i.e. the users who have acquired the right to transmit over specific portions of the spectrum buying the relative license. The most common strategies adopt a hierarchical access structure, distinguishing between *primary* users, or legacy spectrum holders, and *secondary* users, who access the licensed spectrum dynamically, under the constraint of not inducing Quality of Service (QoS) degradations intolerable to the primary users. Within this context, three basic approaches have been considered to allow concurrent communications: *spectrum overlay, underlay* and *interweave.*[1]

[1]There is no strict consensus on some of the basic terminology in cognitive systems [4]. Here we use interweave as in [5] which is sometimes referred to as overlay communications [4].

In overlay systems, as proposed in [7], secondary users allocate part of their power for secondary transmissions and the remainder to assist (relay) the primary transmissions. By exploiting sophisticated coding techniques such as dirty paper coding based on the knowledge of the primary users' message and/or codebook at the cognitive transmitter, these systems offer the possibility of concurrent transmissions without capacity penalties. However, although interesting from an information theoretic perspective, these techniques are difficult to implement as they require noncausal knowledge of the primary signals at the cognitive transmitters.

In underlay systems, the secondary users are also allowed to share resources with the primary users, but without any knowledge about the primary users' signals and under the strict constraint that the spectral density of their transmitted signals fall below the noise floor at the primary receivers. This interference constraint can be met using spread spectrum or ultra-wideband communications from the secondary users. Both transmission techniques do not require the estimation of the electromagnetic environment from secondary users, but they are mostly appropriate for short distance communications, because of the strong constraints imposed on the maximum power radiated by the secondary users.

Conversely, interweave communications, initially envisioned in [6], are based on an *opportunistic* or adaptive usage of the spectrum, as a function of its real utilization. Secondary users are allowed to adapt their power allocation as a function of time and frequency, depending on what they are able to sense and learn from the environment, in a nonintrusive manner. Rather than imposing a severe constraint on their transmit power spectral density, in interweave systems, the secondary users have to figure out *when* and *where* to transmit. Differently from underlay systems, this opportunistic spectrum access requires an opportunity identification phase, through spectrum sensing, followed by an opportunity exploitation mode [4]. For a fascinating motivation and discussion of the signal processing challenges faced in interweave cognitive radio systems, we suggest the interested reader to refer to [2].

In this paper we focus on opportunistic resource allocation techniques in hierarchical cognitive networks, as they seem to be the most suitable for the current spectrum management policies and legacy wireless systems [4]. We are specifically interested in devising the most appropriate form of concurrent communications of cognitive users competing over the physical resources let available from primary users. Looking at opportunistic communication paradigm from a broad signal processing perspective, the secondary users are allowed to transmit over a multi-dimensional space, whose coordinates represent time slots, frequency bins and (possibly) angles, and their goal is to find out the most appropriate transmission strategy, assuming a given power budget at each node, exploring all available degrees of freedom, under the constraint of inducing a limited interference, or no interference at all, at the primary users.

In general, the optimization of the transmission strategies requires the presence of a central node having full knowledge of all the channels and interference structure at every receiver. But this poses a serious implementation problem in terms of scalability and amount of signaling to be exchanged among the nodes. The required extra signaling could, in the end, jeopardize the promise for higher efficiency. To overcome this difficulty, we concentrate on decentralized strategies, where the cognitive users are able to self-enforce the negotiated agreements on the spectrum usage without the intervention of a centralized authority. The philosophy underlying this approach is a *competitive optimality* criterion, as every user aims for the transmission strategy that unilaterally maximizes his own payoff function. The presence of concurrent secondary users competing over the same resources adds dynamics to the system, as every secondary user will dynamically react to the strategies adopted by the other users. The main question is then to establish whether, and under what conditions, the overall system can eventually converge to an equilibrium from which every user is not willing to unilaterally move, as this would determine a performance loss. This form of equilibrium coincides with the well-known concept of Nash Equilibrium (NE) in game theory (see, e.g., [8, 9]). In fact, game theory is the natural tool to devise decentralized strategies allowing the secondary users to find out their best response to any given channel and interference scenario and to derive the conditions for the existence and uniqueness of NE.

Within this context, in this paper, we propose and analyze a totally decentralized approach to design cognitive MIMO transceivers, satisfying a competitive optimality criterion, based on the achievement of Nash equilibria. To take full advantage of all the opportunities offered by wireless communications, we assume a fairly general MIMO setup, where the multiple channels may be frequency channels (as in OFDM systems) [10]-[12], time slots (as in TDMA systems) [10, 11], and/or spatial channels (as in transmit/receive beamforming systems) [13]. Whenever available, multiple antennas at the secondary transmitters could be used, for example, to put nulls in the antenna radiation pattern of secondary transmitters along the directions identifying the primary receivers, thus enabling the share of frequency and time resources with no additional interference. Our initial goal is to provide conditions for the existence and uniqueness of NE points in a game where secondary users compete against each other to maximize their performance, under the constraint on the maximum (or null) interference induced on the primary users. The next step is then to describe low-complex totally distributed techniques able to reach the equilibrium points of the proposed games, with no coordination among the secondary users.

# 2 System Model: Cognitive Radio Networks

We consider a scenario composed by heterogeneous wireless systems (primary and secondary users), as illustrated in Figure 1. The setup may include peer-to-peer links, multiple access, or broadcast channels. The systems coexisting in the network do not have a common goal and do not cooperate with each other. Moreover, no centralized authority is assumed to handle the network access from secondary users. Thus, the secondary users are allowed, in principle, to compete for the same physical resources, e.g., time, frequency, and space. We are interested in finding the optimal transmission strategy for the secondary users, using a decentralized approach. A fairly general system model to describe the signals received by the secondary users is the Gaussian *vector* interference channel:

$$\mathbf{y}_q = \mathbf{H}_{qq}\mathbf{x}_q + \sum_{r\neq q} \mathbf{H}_{rq}\mathbf{x}_r + \mathbf{n}_q, \tag{1}$$

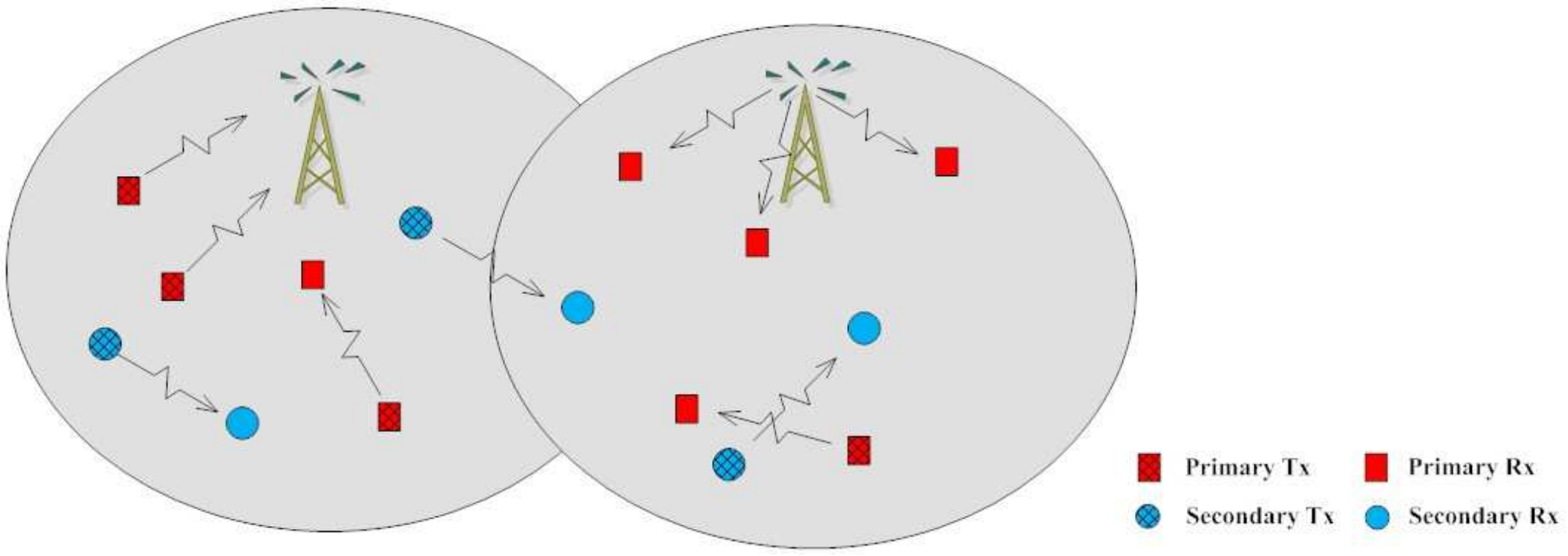

Figure 1: Hierarchical cognitive radio network with primary and secondary users.

where $\mathbf{x}_q$ is the $n_{T_q}$-dimensional block of data transmitted by source $q$, $\mathbf{H}_{qq}$ is the $n_{R_q} \times n_{T_q}$ (complex) channel matrix between the $q$-th transmitter and its intended receiver, $\mathbf{H}_{rq}$ is the $n_{R_q} \times n_{T_r}$ cross-channel matrix between source $r$ and destination $q$, $\mathbf{y}_q$ is the $n_{R_q}$-dimensional vector received by destination $q$, and $\mathbf{n}_q$ is the $n_{R_q}$-dimensional noise plus interference vector. The first term in the right-hand side of (1) is the useful signal for link $q$, the second and third terms represent the Multi-User Interference (MUI) received by secondary user $q$ and caused from the other secondary users and the primary users, respectively. The vector $\mathbf{n}_q$ is assumed to be zero-mean circularly symmetric complex Gaussian with arbitrary (nonsingular) covariance matrix $\mathbf{R}_{n_q}$. For the sake of simplicity and lack of space, we consider here only the case where the channel matrices $\mathbf{H}_{qq}$ are square nonsingular. We assume that each receiver is able to estimate the channel from its intended transmitter and the overall MUI covariance matrix (alternatively, to make short term predictions, with negligible error).[2] The receiver sends then this information back to the transmitter through a low bit rate (error-free) feedback channel, to allow the transmitter to compute the optimal transmission strategy over its own link.

The model in (1) represents a fairly general MIMO setup, describing multiuser transmissions over multiple channels, which may represent frequency channels (as in OFDM systems) [10]-[12], time slots (as in TDMA systems) [10, 11], or spatial channels (as in transmit/receive beamforming systems) [13]. Differently from traditional static or centralized spectrum assignment, the cognitive radio paradigm enables secondary users to transmit with overlapping spectrum and/or coverage with primary users, provided that the degradation induced on the primary users' performance is null or tolerable. How to impose interference constraints on secondary users is a complex and open regulatory issue [2, 4]. Roughly speaking, restrictive constraints may marginalize the potential gains offered by the dynamic resource assignment mechanism, whereas loose constraints may affect the compatibility with legacy systems. Both deterministic and probabilistic interference constraints have been suggested in the literature [1, 2, 4, 15], namely: the maximum MUI interference power level perceived by any active primary user (the so-called *interference temperature limit*) [1, 2] and the maximum probability that the MUI interference level at each primary user's receiver may exceed a prescribed threshold [4, 15]. In the presence of sensing errors, the access to channels identified as idle should also depend on the goodness of the channel estimation. As shown in [17], in this case the optimal strategy is probabilistic, with an probability depending on both the false alarm and miss probabilities.

[2] How to obtain both channel-state information and MUI covariance matrix estimation goes beyond the scope of this paper; the interested reader may refer to, e.g., [2, 4], where classical signal processing estimation techniques are properly modified to be successfully applied in a cognitive radio environment.

In this paper we are primarily interested in analyzing the contention among the secondary users over a multiuser channel where there are primary users as well. To limit the complexity of the problem, in the effort to find out distributed techniques guaranteed to converge to NE points, we restrict our analysis to consider only deterministic interference constraints, albeit expressed in a very general form. In particular, we envisage the use of the following possible interference constraints (see also Figure 2):

**Co.1** *Maximum transmit power for each transmitter*:

$$\mathcal{E}\left\{\|\mathbf{x}_q\|_2^2\right\} = \mathsf{Tr}\left(\mathbf{Q}_q\right) \leq P_q, \tag{2}$$

where $\mathbf{Q}_q$ denotes the covariance matrix of the symbols transmitted by user $q$ and $P_q$ is the transmit power in units of energy per transmission.

**Co.2** *Null constraints*:

$$\mathbf{U}_q^H \mathbf{Q}_q = \mathbf{0}, \tag{3}$$

where $\mathbf{U}_q$ is a strict tall matrix (to avoid the trivial solution $\mathbf{Q}_q = \mathbf{0}$), whose columns represent the spatial and/or the frequency "directions" along with user $q$ is not allowed to transmit. We assume, without loss of generality (w.l.o.g.), that each matrix $\mathbf{U}_q$ is full-column rank.

**Co.3** *Soft shaping constraints*:

$$\mathsf{Tr}\left(\mathbf{G}_q^H \mathbf{Q}_q \mathbf{G}_q\right) \leq P_q^{\text{ave}}, \tag{4}$$

where the matrices $\mathbf{G}_q$ are such that their range space identifies the subspace where the interference level should be kept under the required threshold.[3]

**Co.4** *Peak power constraints*: the average peak power of each user $q$ can be controlled by constraining the maximum eigenvalue [denoted by $\lambda_{\max}(\cdot)$] of the transmit covariance matrix along the directions spanned by the column space of $\mathbf{G}_q$:

$$\lambda_{\max}\left(\mathbf{G}_q^H \mathbf{Q}_q \mathbf{G}_q\right) \leq P_q^{\text{peak}}, \tag{5}$$

where $P_q^{\text{peak}}$ is the maximum peak power that can be transmitted along the spatial and/or the frequency directions spanned by the column space of $\mathbf{G}_q$.

[3]The interference temperature limit constraint [2] is given by the aggregated interference induced by all secondary users. In this paper, we assume that the primary user imposing the soft constraint, has already computed the maximum tolerable interference power $P_q^{\text{ave}}$ for each secondary user. The power limit $P_q^{\text{ave}}$ can also be the result of a negotiation or opportunistic based procedure between primary users (or regulatory agencies) and secondary users.

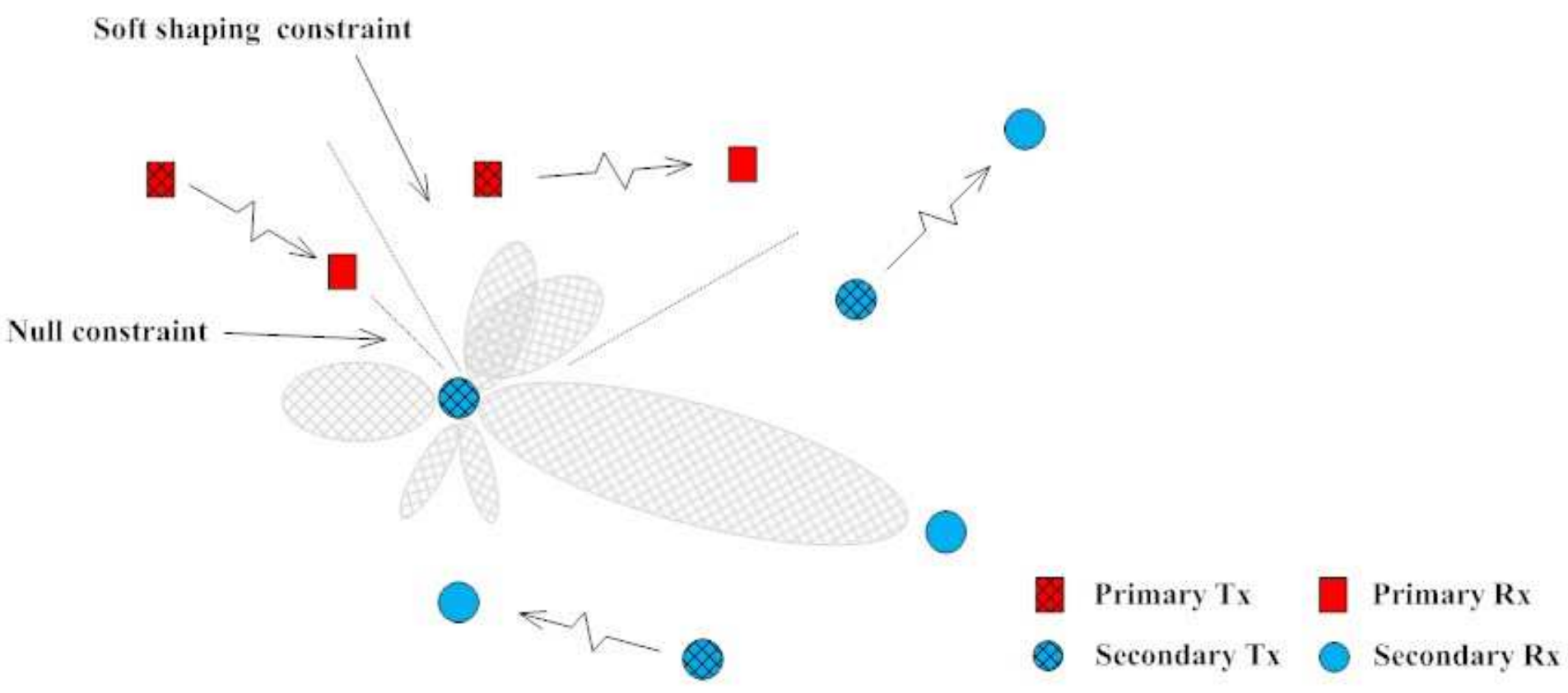

Figure 2: Example of null/soft shaping constraints.

The structure of the null constraints in (3) is a very general form to express the strict limitation imposed on secondary users to prevent them from transmitting over the subchannels occupied by the primary users. These subchannels are modeled as vectors belonging to the subspace spanned by the columns of each matrix $\mathbf{U}_q$. This form includes, as particular cases, the imposition of nulls over: 1) the frequency bands occupied by the primary receivers; 2) the time slots occupied by the primary users; 3) the angular directions identifying the primary receivers as observed from the secondary transmitters. In the first case, the subspace is spanned by a set of IFFT vectors, in the second case by a set of canonical vectors, and in the third case by the set of steering vectors representing the directions of the primary receivers as observed from the secondary transmitters. It is worth emphasizing that the structure of the null constraints in (3) is much more general than the three cases mentioned above, as it can incorporate any combination of the frequency, time and space coordinates.

The use of the spatial domain can greatly improve the capabilities of cognitive users, as it allows them to transmit over the same frequency band but without interfering. This is possible if the secondary transmitters have an antenna array and use a beamforming that puts nulls over the directions identifying the primary receivers. Of course, this requires the identification of the primary receivers, a task that is much more demanding than the detection of primary transmitters [4]. As an example, there are some recent works showing that, in the application of CR over the spectrum allocated to commercial TV, one might exploit the local oscillator leakage power emitted by the RF front end of the TV receiver to locate the receivers [18]. Of course, in such a case, the detection range is quite short and this calls for a deployment of sensors very close to the potential receivers. A different scenario pertains to cellular systems. In such a case, the mobile users might be rather hard to locate and track. However, the base stations are relatively easier to identify. Hence, in a cellular system operating in a time-division duplexing (TDD) mode, the secondary users could exploit the time slot allocated for the uplink channel and put a null in the direction of the base stations. This would avoid any interference towards the cellular system users, without the need of tracking the mobile users.

The soft shaping constraints expressed in (4) and (5) represent a constraint on the total average and peak average power radiated (projected) along the directions spanned by the column space of matrix $\mathbf{G}_q$. They are a relaxed form of (3) and can be used to keep the portion of the interference temperature generated by each secondary user $q$ under the desired value. In fact, under (4)-(5), the secondary users are allowed to transmit over some subchannels occupied by the primary users, but only provided that the interference that they generate falls below a prescribed threshold. For example, in a MIMO setup, the matrix $\mathbf{G}_q$ in (4) would contain, in its columns, the steering vectors identifying the directions of the primary receivers.

Within the assumptions made above, invoking the capacity expression for the single user Gaussian MIMO channel−achievable using random Gaussian codes by all the users−the maximum information rate on link $q$ for a given set of users' covariance matrices $\mathbf{Q}_1, \ldots, \mathbf{Q}_Q$, is [19]

$$R_q(\mathbf{Q}_q, \mathbf{Q}_{-q}) = \log\det\left(\mathbf{I} + \mathbf{H}_{qq}^H \mathbf{R}_{-q}^{-1} \mathbf{H}_{qq} \mathbf{Q}_q\right) \tag{6}$$

where

$$\mathbf{R}_{-q} \triangleq \mathbf{R}_{n_q} + \sum_{r \neq q} \mathbf{H}_{rq} \mathbf{Q}_r \mathbf{H}_{rq}^H \tag{7}$$

is the MUI plus noise covariance matrix observed by user $q$ and $\mathbf{Q}_{-q} \triangleq (\mathbf{Q}_r)_{r \neq q}$ is the set of all the users' covariance matrices, except the $q$-th one. Observe that $\mathbf{R}_{-q}$ depends on the strategies $\mathbf{Q}_{-q}$ of the other players.

## 3 Resource Sharing among Secondary Users based on Game Theory

Given the multiuser nature of the scenario described above, the design of the optimal transmission strategies of secondary users would require a multiobjective formulation of the optimization problem, as the information rate achieved on each secondary user's link constitutes a different single objective function. The globally optimal solutions of such a problem−the Pareto optimal surface of the multiobjective problem−would define the largest rate region achievable by secondary users, given the power constraints **Co.1**-**Co.4**: the rate vector profile $\mathbf{R}(\mathbf{Q}^\star) \triangleq [R_1(\mathbf{Q}^\star), \ldots, R_Q(\mathbf{Q}^\star)]$ is Pareto optimal if there exists no other rate profile $\mathbf{R}(\mathbf{Q})$ that dominates $\mathbf{R}(\mathbf{Q}^\star)$ component-wise, i.e., $\mathbf{R}(\mathbf{Q}^\star) \geq \mathbf{R}(\mathbf{Q})$, for all feasible $\mathbf{Q}$'s, where at least one inequality is strict.

Unfortunately, the computation of the rate region is analytically intractable and thus not applicable in a cognitive radio scenario, since every scalar/multiobjective optimization problem involving the rates of secondary users in (6) is not convex (implied from the fact that the rates $R_q(\mathbf{Q})$ are nonconcave functions of the covariance matrices $\mathbf{Q}$). Furthermore, even in the simpler case of transmissions over SISO parallel channels, the network utility maximization (NUM) problem based on the rates functions (6) has been proved in [24] to be a strongly NP-hard problem, under various practical settings as well as different choices of the system utility function (e.g., sum-rate, weighted sum-rate, geometric rate-mean). Roughly speaking, this means that there is no hope to obtain an algorithm, even centralized, that can efficiently compute the exact globally optimal solution. Although in theory, the rate region could be still found by an exhaustive search through all possible feasible covariance matrices, the computational complexity of this approach is prohibitively high, given the large number of variables and users involved in the optimization. The situation is particularly critical in CR systems, where the cognitive users sense a very large spectrum. Consequently, *suboptimal* algorithms have been proposed in the literature to solve special cases of the proposed optimization [20]-[23], most of them dealing with the maximization of the (weighted) sum-rate in SISO frequency-selective interference channels (obtained from our general model when the channel matrices are diagonal, the covariance matrices reduce to the power allocation vectors, and the null/soft shaping constraints are removed) [20, 21]. Due to the nonconvex nature of the problem, these algorithms either lack global convergence or may converge to poor spectrum sharing strategies.

Furthermore, even if one decides to employ a suboptimal method, e.g., [20]-[23], the algorithms are not suitable for CR systems as they are centralized and thus cannot be implemented in a distributed way. These techniques require a central authority (or node in the network) with knowledge of the (direct and cross-) channels to compute all the transmission strategies for the different nodes and then to broadcast the solution. This scheme would clearly pose a serious implementation problem in terms of scalability of the network and amount of signaling to be exchanged among the nodes, which makes such an approach not appealing in the scenario considered in this paper.

To overcome the above difficulties and reach a better trade-off between performance and complexity, we shift our focus to a different notion of optimality: the competitive optimality criterion; which motivates a game theoretical formulation of the system design. Using the concept of NE as the competitive optimality criterion, the resource allocation problem among secondary users is then cast as a strategic noncooperative game, in which the players are the secondary users and the payoff functions are the information rates on each link: Each secondary user $q$ competes against the others by choosing the transmit covariance matrix $\mathbf{Q}_q$ (i.e., his strategy) that maximizes his own information rate $R_q(\mathbf{Q}_q, \mathbf{Q}_{-q})$ in (6), given constraints imposed by the presence of the primary users, besides the usual constraint on transmit power. A NE of the game is reached when each user, given the strategy profiles of the others, does not get any rate increase by unilaterally changing his own strategy. The first question to answer under such framework is whether such an overall dynamical system can eventually converge to an equilibrium point, while preserving the QoS of primary users. The second basic issue is if the optimal strategies to be adopted by each user can be computed in a totally decentralized way. We address both questions in the forthcoming sections.

For the sake of simplicity, we start considering only constraints **Co.1** and **Co.2**. These constraints are suitable to model interweave communications among secondary users where, in general, there are restrictions on when and where they may transmit (this can be done using the null constraints **Co.2**). Then, we allow underlay and interweave communications simultaneously, by including in the optimization also interference constraints **Co.3** and **Co.4**.

### 3.1 Rate maximization game with null constraints

Given the rate functions in (6) and constraints **Co.1-Co.2**, the rate maximization game is formally defined as:

$$(\mathcal{G}_1): \quad \begin{array}{ll} \underset{\mathbf{Q}_q \succeq \mathbf{0}}{\text{maximize}} & R_q(\mathbf{Q}_q, \mathbf{Q}_{-q}) \\ \text{subject to} & \mathsf{Tr}\left(\mathbf{Q}_q\right) \leq P_q, \quad \mathbf{U}_q^H \mathbf{Q}_q = \mathbf{0} \end{array} \qquad \forall q = 1, \cdots, Q, \tag{8}$$

where $Q$ is the number of players (the secondary users) and $R_q(\mathbf{Q}_q, \mathbf{Q}_{-q})$ is the payoff function of player $q$, defined in (6). Without the null constraints, the solution of each optimization problem in (8) would lead to the well-known MIMO waterfilling solution [19]. The presence of the null constraints modifies the problem and the solution for each user is not necessarily a waterfilling anymore. Nevertheless, we show now that introducing a proper projection matrix the solutions of (8) can still be efficiently computed via a waterfilling-like expression. To this end, we rewrite game $\mathcal{G}_1$ in a more convenient form as detailed next.

Introducing the projection matrix $\mathbf{P}_{\mathcal{R}(\mathbf{U}_q)^\perp} = \mathbf{I} - \mathbf{U}_q(\mathbf{U}_q^H\mathbf{U}_q)^{-1}\mathbf{U}_q^H$ (the orthogonal projection onto $\mathcal{R}(\mathbf{U}_q)^\perp$, where $\mathcal{R}(\cdot)$ is the range space operator), it follows from the constraint $\mathbf{U}_q^H\mathbf{Q}_q = \mathbf{0}$ that any optimal $\mathbf{Q}_q$ in (8) will always satisfy:

$$\mathbf{Q}_q = \mathbf{P}_{\mathcal{R}(\mathbf{U}_q)^\perp}\mathbf{Q}_q\mathbf{P}_{\mathcal{R}(\mathbf{U}_q)^\perp}. \tag{9}$$

The game $\mathcal{G}_1$ can then be equivalently rewritten as:

$$\begin{array}{ll} \underset{\mathbf{Q}_q \succeq \mathbf{0}}{\text{maximize}} & \log\det\left(\mathbf{I} + \tilde{\mathbf{H}}_{qq}^H\tilde{\mathbf{R}}_{-q}^{-1}\tilde{\mathbf{H}}_{qq}\mathbf{Q}_q\right) \\ \text{subject to} & \mathsf{Tr}\left(\mathbf{Q}_q\right) \leq P_q \\ & \mathbf{Q}_q = \mathbf{P}_{\mathcal{R}(\mathbf{U}_q)^\perp}\mathbf{Q}_q\mathbf{P}_{\mathcal{R}(\mathbf{U}_q)^\perp} \end{array} \qquad \forall q = 1, \cdots, Q, \tag{10}$$

where each $\tilde{\mathbf{H}}_{rq} \triangleq \mathbf{H}_{rq}\mathbf{P}_{\mathcal{R}(\mathbf{U}_r)^\perp}$ is a modified channel and $\tilde{\mathbf{R}}_{-q} \triangleq \mathbf{R}_{n_q} + \sum_{r\neq q}\tilde{\mathbf{H}}_{rq}\mathbf{Q}_r\tilde{\mathbf{H}}_{rq}^H$. At this point, the problem can be further simplified by noting that the constraint $\mathbf{Q}_q = \mathbf{P}_{\mathcal{R}(\mathbf{U}_q^\perp)}\,\mathbf{Q}_q\mathbf{P}_{\mathcal{R}(\mathbf{U}_q^\perp)}$ in (10) is redundant. The final formulation then becomes:

$$\begin{array}{ll} \underset{\mathbf{Q}_q \succeq \mathbf{0}}{\text{maximize}} & \log\det\left(\mathbf{I} + \tilde{\mathbf{H}}_{qq}^H\tilde{\mathbf{R}}_{-q}^{-1}\tilde{\mathbf{H}}_{qq}\mathbf{Q}_q\right) \\ \text{subject to} & \mathsf{Tr}(\mathbf{Q}_q) \leq P_q \end{array} \qquad \forall q = 1, \cdots, Q. \tag{11}$$

This is due to the fact that, for any user $q$, any optimal solution $\mathbf{Q}_q^\star$ in (11)$-$the MIMO waterfilling solution [13]$-$will be orthogonal to the null space of $\tilde{\mathbf{H}}_{qq}$, whatever $\tilde{\mathbf{R}}_{-q}$ is, implying $\mathbf{Q}_q^\star = \mathbf{P}_{\mathcal{R}(\mathbf{U}_q)^\perp}\mathbf{Q}_q^\star\mathbf{P}_{\mathcal{R}(\mathbf{U}_q)^\perp}$. Building on the equivalence of (8) and (11), we can apply the results in [13] to the game in (11) and derive the structure of the Nash equilibria of game $\mathcal{G}_1$, as detailed next.

**Nash equilibria of game $\mathcal{G}_1$:** Game $\mathcal{G}_1$ always admits a NE, for any set of channel matrices, transmit power of the users, and null constraints, since it is a concave game (the payoff of each player is a concave function in his own strategy and each admissible strategy set is convex and compact) [13]. Moreover, it follows from (11) that all the Nash equilibria of $\mathcal{G}_1$ satisfy the following set of nonlinear matrix-value fixed-point equations [13]:

$$\mathbf{Q}_q^\star = \tilde{\mathbf{WF}}_q\left(\tilde{\mathbf{H}}_{qq}^H\mathbf{R}_{-q}^{-1}(\mathbf{Q}_{-q}^\star)\tilde{\mathbf{H}}_{qq}\right) \triangleq \mathbf{W}_q^\star\,\mathrm{Diag}\left(\mathbf{p}_q^\star\right)\mathbf{W}_q^{\star H}, \quad \forall q = 1, \cdots, Q, \tag{12}$$

where we made explicit the dependence of $\mathbf{R}_{-q}$ on $\mathbf{Q}^{\star}_{-q}$ as $\mathbf{R}_{-q}(\mathbf{Q}^{\star}_{-q})$; the $\mathbf{W}^{\star}_{q} = \mathbf{W}_{q}(\mathbf{Q}^{\star}_{-q})$ is the semi-unitary matrix with columns equal to the eigenvectors of matrix $\tilde{\mathbf{H}}^{H}_{qq}\mathbf{R}^{-1}_{-q}(\mathbf{Q}^{\star}_{-q})\tilde{\mathbf{H}}_{qq}$ corresponding to the positive eigenvalues $\lambda^{\star}_{q,k} = \lambda_{q,k}(\mathbf{Q}^{\star}_{-q})$, with $\mathbf{R}_{-q}(\mathbf{Q}_{-q})$ defined in (7); and the power allocation $\mathbf{p}^{\star}_{q} = \mathbf{p}_{q}(\mathbf{Q}^{\star}_{-q})$ satisfies the following simultaneous waterfilling equation: for all $k$ and $q$,

$$p^{\star}_{q}(k) = \left(\mu_q - \frac{1}{\lambda^{\star}_{q,k}}\right)^{+}, \tag{13}$$

with $(x)^{+} \triangleq \max(0, x)$ and $\mu_q$ chosen to satisfy the power constraint $\sum_k p^{\star}_{q}(k) = P_q$.

Interestingly, the solution (12) shows that the null constraints in the transmissions of secondary users can be handled without affecting the computational complexity: The optimal transmission strategy of each user $q$ can be efficiently computed via a MIMO waterfilling solution, provided that the original channel matrix $\mathbf{H}_{qq}$ is replaced by $\tilde{\mathbf{H}}_{qq}$.

This result has an intuitive interpretation: To guarantee that each user $q$ does not transmit over a given subspace (spanned by the columns of $\mathbf{U}_q$), *whichever* the strategies of the other users are, while maximizing his information rate, one only needs to induce in the channel matrix $\mathbf{H}_{qq}$ a null space that coincides with the subspace where the transmission is not allowed. This is precisely what is done by introducing the modified channel $\tilde{\mathbf{H}}_{qq}$.

The waterfilling-like structure of the Nash equilibria as given in (12) along with the interpretation of the MIMO watefilling solution as a matrix projection onto a proper convex set as given in [13] play a key role in studying the uniqueness of the NE and in deriving conditions for the convergence of the distributed algorithms described in Section 4. The analysis of the uniqueness of the NE goes beyond the scope of this paper and it is addressed in [14]. What is important to remark here is that, as expected, the conditions guaranteeing the uniqueness of the NE impose a constraint on the maximum level of MUI generated by secondary users that may be tolerated in the network. But, interestingly, the uniqueness of the equilibrium is not affected by the interference generated by the primary users.

### 3.2 Rate maximization game with null constraints via virtual noise shaping

In this section, we show that an alternative approach to impose null constraints **Co.2** on the transmissions of secondary users passes through the introduction of virtual interferers. The idea behind this alternative approach can be easily understood if one considers the transmission over SISO frequency-selective channels, where all the channel matrices have the same eigenvectors (the FFT vectors): to avoid the use of a given subchannel, it is sufficient to introduce a "virtual" noise with sufficiently high power over that subchannel. The same idea cannot be directly applied to the MIMO case, as arbitrary MIMO channel matrices have different right/left singular vectors from each other. Nevertheless, we show how to design the covariance matrix of the virtual noise (to be added to the noise covariance matrix of each secondary receiver), so that the all the Nash equilibria of the game satisfy the null constraint **Co.2** along the specified directions.

Let us consider the following strategic noncooperative game:

$$(\mathcal{G}_\alpha): \qquad \begin{array}{ll} \underset{\mathbf{Q}_q\succeq\mathbf{0}}{\text{maximize}} & \log\det\left(\mathbf{I}+\mathbf{H}_{qq}^H\mathbf{R}_{-q,\alpha}^{-1}\mathbf{H}_{qq}\mathbf{Q}_q\right) \\ \text{subject to} & \mathsf{Tr}\left(\mathbf{Q}_q\right)\leq P_q \end{array} \qquad \forall q=1,\cdots,Q, \tag{14}$$

where

$$\mathbf{R}_{-q,\alpha}\triangleq\mathbf{R}_{-q}+\alpha\hat{\mathbf{U}}_q\hat{\mathbf{U}}_q^H=\mathbf{R}_{n_q}+\sum_{r\neq q}\mathbf{H}_{rq}\mathbf{Q}_r\mathbf{H}_{rq}^H+\alpha\hat{\mathbf{U}}_q\hat{\mathbf{U}}_q^H, \tag{15}$$

denotes the MUI-plus-noise covariance matrix observed by secondary user $q$, plus the covariance matrix $\alpha\hat{\mathbf{U}}_q\hat{\mathbf{U}}_q^H$ of the virtual interference along $\mathcal{R}(\hat{\mathbf{U}}_q)$, where $\hat{\mathbf{U}}_q$ is a tall matrix and $\alpha$ is a positive constant. Our interest is on deriving the asymptotic properties of the solutions of $\mathcal{G}_\alpha$, as $\alpha\to+\infty$. To this end, we introduce the following intermediate definitions first. For each $q$, define the tall matrix $\hat{\mathbf{U}}_q^\perp$ such that $\mathcal{R}(\hat{\mathbf{U}}_q^\perp)=\mathcal{R}(\hat{\mathbf{U}}_q)^\perp$, and the modified channel matrices

$$\hat{\mathbf{H}}_{rq}=\hat{\mathbf{U}}_q^{\perp^H}\mathbf{H}_{rq} \qquad \forall r,\, q=1,\cdots,Q. \tag{16}$$

We then introduce the auxiliary game $\mathcal{G}_\infty$, defined as:

$$(\mathcal{G}_\infty): \qquad \begin{array}{ll} \underset{\mathbf{Q}_q\succeq\mathbf{0}}{\text{maximize}} & \log\det\left(\mathbf{I}+\hat{\mathbf{H}}_{qq}^H\hat{\mathbf{R}}_{-q}^{-1}\hat{\mathbf{H}}_{qq}\mathbf{Q}_q\right) \\ \text{subject to} & \mathsf{Tr}\left(\mathbf{Q}_q\right)\leq P_q \end{array} \qquad \forall q=1,\cdots,Q, \tag{17}$$

where

$$\hat{\mathbf{R}}_{-q}\triangleq\hat{\mathbf{U}}_q^{\perp H}\mathbf{R}_{n_q}\hat{\mathbf{U}}_q^\perp+\sum_{r\neq q}\hat{\mathbf{H}}_{rq}\mathbf{Q}_r\hat{\mathbf{H}}_{rq}^H. \tag{18}$$

It can be shown that games $\mathcal{G}_\alpha$ and $\mathcal{G}_\infty$ are asymptotically equivalent in the sense specified next.

**Nash equilibria of games $\mathcal{G}_\alpha$ and $\mathcal{G}_\infty$:** Games $\mathcal{G}_\alpha$ and $\mathcal{G}_\infty$ always admit a NE, for any set of channel matrices, power constraints, and $\alpha>0$. Moreover, under mild conditions guaranteeing the uniqueness of the NE of both games (denoted by $\mathbf{Q}_\alpha^\star$ and $\mathbf{Q}_\infty^\star$, respectively), we have:

$$\lim_{\alpha\to\infty}\mathbf{Q}_\alpha^\star=\mathbf{Q}_\infty^\star, \tag{19}$$

i.e., the NE of $\mathcal{G}_\alpha$ asymptotically coincides with that of $\mathcal{G}_\infty$.

Observe that, similarly to game $\mathcal{G}_1$, also in games $\mathcal{G}_\alpha$ and $\mathcal{G}_\infty$, the best-response of each player can be efficiently computed via MIMO waterfilling-like solutions, and the Nash equilibria of both games satisfy a simultaneous waterfilling equation.

Using (19), one can derive the asymptotic properties of the (unique) NE of game $\mathcal{G}_\alpha$ as $\alpha\to\infty$, through the properties of the equilibrium $\mathbf{Q}_\infty^\star$ of $\mathcal{G}_\infty$. Following a similar approach as in Section 3.1, one can show that each $\mathbf{Q}_{q,\infty}^\star$ satisfies the following condition

$$\mathbf{U}_q^H\mathbf{Q}_{q,\infty}^\star=\mathbf{0}, \qquad \text{with} \qquad \mathbf{U}_q\triangleq\mathbf{H}_{qq}^{-1}\hat{\mathbf{U}}_q. \tag{20}$$

Condition (20) provides, for each user $q$, the desired relationship between the directions of the virtual noise to be introduced in the noise covariance matrix of the user (see (18))$-$the matrix $\hat{\mathbf{U}}_q-$and the real directions along with user $q$ will not allocate any power, i.e., the matrix $\mathbf{U}_q$. It turns out that if user $q$ is not allowed to allocate power along $\mathbf{U}_q$, it is sufficient to choose in (18) $\hat{\mathbf{U}}_q \triangleq \mathbf{H}_{qq}\mathbf{U}_q$.

Since the existence and uniqueness of the NE of game $\mathcal{G}_\alpha$ do not depend on $\alpha$, the (unique) NE of $\mathcal{G}_\alpha$ (that in general will depend on the value of $\alpha$) can be reached using the asynchronous algorithms described in Section 4, irrespective of the value of $\alpha$. Thus, for sufficiently large values of $\alpha$, the NE of $\mathcal{G}_\alpha$ tends to satisfy condition (20); which provides an alternative way to impose constraint **Co.2**.

### 3.3 Rate maximization game with soft and null constraints

We focus now on the rate maximization in the presence of both null and *soft shaping* constraints. The resulting game can be formulated as follows:

$$(\mathcal{G}_2): \quad \begin{array}{ll} \underset{\mathbf{Q}_q \succeq \mathbf{0}}{\text{maximize}} & R_q(\mathbf{Q}_q, \mathbf{Q}_{-q}) \\ \text{subject to} & \mathsf{Tr}\left(\mathbf{G}_q^H \mathbf{Q}_q \mathbf{G}_q\right) \leq P_q^{\text{ave}} \\ & \lambda_{\max}\left(\mathbf{G}_q^H \mathbf{Q}_q \mathbf{G}_q\right) \leq P_q^{\text{peak}} \\ & \mathbf{U}_q^H \mathbf{Q}_q = \mathbf{0} \end{array} \quad \forall q = 1, \cdots, Q. \tag{21}$$

We assume w.l.o.g. that each $\mathbf{G}_q$ is a full-row rank matrix, so that the soft shaping constraint in (21) imposes a constraint on the average transmit power radiated by user $q$ in the *whole* space.

The soft constraints in (21) are the result of a constraint on the overall interference temperature limit imposed by the primary users [2]. Typically, the most stringent conditions between the power constraints **Co.1** and **Co.3** is the soft shaping constraint **Co.3**. This motivates the absence in (21) of the power constraint **Co.1**, although it could also be considered.

**Nash equilibria of game $\mathcal{G}_2$:** We can derive the structure of the Nash equilibria of game $\mathcal{G}_2$, similarly to what we did for game $\mathcal{G}_1$. For each $q \in \Omega$, define the tall matrix $\overline{\mathbf{U}}_q \triangleq \mathbf{G}_q^\sharp \mathbf{U}_q$, where $\mathbf{G}_q^\sharp$ denotes the Monroe-Penrose pseudoinverse of $\mathbf{G}_q$ [25], introduce the projection matrix $\mathbf{P}_{\mathcal{R}(\overline{\mathbf{U}}_q)^\perp} = \mathbf{I} - \overline{\mathbf{U}}_q(\overline{\mathbf{U}}_q^H \overline{\mathbf{U}}_q)^{-1}$ $\overline{\mathbf{U}}_q^H$ (the orthogonal projection onto $\mathcal{R}(\overline{\mathbf{U}}_q)^\perp$) and the modified channel matrices

$$\overline{\mathbf{H}}_{rq} = \mathbf{H}_{rq} \mathbf{G}_r^{\sharp^H} \mathbf{P}_{\mathcal{R}(\overline{\mathbf{U}}_r)^\perp}, \quad r, q = 1, \cdots, Q. \tag{22}$$

Using the above definition, we can now characterize the Nash equilibria of game $\mathcal{G}_2$, as shown next.

The game $\mathcal{G}_2$ admits a NE, for any set of channel matrices and null/soft shaping constraints. Moreover, every NE satisfies the following set of nonlinear matrix-value fixed-point equations:

$$\begin{array}{rcl} \mathbf{Q}_q^\star & = & \mathbf{G}_q^{\sharp^H} \overline{\mathbf{WF}}_q \left( \overline{\mathbf{H}}_{qq}^H \mathbf{R}_{-q}^{-1}(\mathbf{Q}_{-q}^\star) \overline{\mathbf{H}}_{qq} \right) \mathbf{G}_q^\sharp \\ & \triangleq & \mathbf{G}_q^{\sharp^H} \mathbf{V}_q^\star \,\text{diag}\left(\mathbf{p}_q^\star\right) \mathbf{V}_q^{\star H} \mathbf{G}_q^\sharp \end{array} \quad \forall q = 1, \cdots, Q, \tag{23}$$

where $\mathbf{V}_q^\star = \mathbf{V}_q(\mathbf{Q}_{-q}^\star)$ is the semi-unitary matrix with columns equal to the eigenvectors of matrix $\overline{\mathbf{H}}_{qq}^H \mathbf{R}_{-q}^{-1}(\mathbf{Q}_{-q}^\star)\overline{\mathbf{H}}_{qq}$, with $\mathbf{R}_{-q}(\mathbf{Q}_{-q})$ defined in (7), corresponding to the $\bar{L}_q = \mathsf{rank}(\overline{\mathbf{H}}_{qq})$ positive eigenvalues $\lambda_{q,k}^\star = \lambda_{q,k}(\mathbf{Q}_{-q}^\star)$, and the power allocation $\mathbf{p}_q^\star = \mathbf{p}_q(\mathbf{Q}_{-q}^\star)$ satisfies the following simultaneous waterfilling equation: for all $k$ and $q$,

$$p_q^\star(k) = \begin{cases} \left[\mu_q - \frac{1}{\lambda_{q,k}^\star}\right]_0^{P_q^{\text{peak}}}, & \text{if } P_q^{\text{peak}}\bar{L}_q > P_q^{\text{ave}}, \\ P_q^{\text{peak}}, & \text{otherwise}, \end{cases} \tag{24}$$

where $[\cdot]_0^{P_q^{\text{peak}}}$ denotes the Euclidean projection onto the interval $[0,\ P_q^{\text{peak}}]$ and $\mu_q$ is chosen to satisfy the power constraint $\sum_k p_q^\star(k) = P_q^{\text{ave}}$ (see, e.g., [26] for practical algorithms to compute such a $\mu_q$).

The structure of the Nash equilibria in (23) states that the optimal transmission strategy of each user leads to a diagonalizing transmission with a proper power allocation, after pre/post multiplication of the waterfilling solution by matrix $\mathbf{G}_q^\sharp$. Similarly to $\mathcal{G}_1$, the conditions for the uniqueness of the NE of game $\mathcal{G}_2$ can be obtained, building on the interpretation of the waterfilling solutions in (23) as matrix projection [13]. As expected, the NE of the game is unique, provided that the interference generated by secondary users is not too high.

# 4 MIMO Asynchronous Iterative Waterfilling Algorithm

In Section 3 we have shown that the optimal resource allocation among secondary users in hierarchical cognitive networks corresponds to an equilibrium of the system, where all the users have maximized their own rates, without hampering the communications of primary users. Since there is no reason to expect a system to be initially at the equilibrium, the fundamental problem becomes to find a procedure that reaches such an equilibrium from non-equilibrium states. In this section, we focus on algorithms that converge to these equilibria. Since we are interested in a decentralized implementation, where no signaling among secondary and primary users is allowed, we consider only totally distributed iterative algorithms, where each user acts independently of the others to optimize his own transmission strategy while perceiving the other active users as interference

More specifically, to reach the Nash equilibria of the games introduced in the previous section, we propose a fairly general distributed and asynchronous iterative algorithm, called asynchronous Iterative WaterFilling Algorithm (IWFA). In this algorithm, all secondary users maximize their own rate [via the single user MIMO waterfilling solution (12) for game $\mathcal{G}_1$, (23) for game $\mathcal{G}_2$, and the classical MIMO waterfilling solution for games $\mathcal{G}_\alpha$ and $\mathcal{G}_\infty$] in a *totally asynchronous* way, while keeping the temperature noise levels in the licensed bands under the required threshold [2]. According to the asynchronous updating schedule, some users are allowed to update their strategy more frequently than the others, and they might even perform these updates using *outdated* information on the interference caused by the others.

Before introducing the proposed asynchronous MIMO IWFA, we need the following preliminary definitions. We assume, without loss of generality, that the set of times at which one or more users update their strategies is the discrete set $\mathcal{T} = \mathbb{N}_+ = \{0, 1, 2, \ldots\}$. Let $\mathbf{Q}_q^{(n)}$ denote the covariance matrix of the vector signal transmitted by user $q$ at the $n$-th iteration, and let $\mathcal{T}_q \subseteq \mathcal{T}$ denote the set of times $n$ at which $\mathbf{Q}_q^{(n)}$ is updated (thus, at time $n \notin \mathcal{T}_q$, $\mathbf{Q}_q^{(n)}$ is left unchanged). Let $\tau_r^q(n)$ denote the most recent time at which the interference from user $r$ is perceived by user $q$ at the $n$-th iteration (observe that $\tau_r^q(n)$ satisfies $0 \leq \tau_r^q(n) \leq n$). Hence, if user $q$ updates his own covariance matrix at the $n$-th iteration, then he chooses his optimal $\mathbf{Q}_q^{(n)}$, according to (12) for game $\mathcal{G}_1$ and (23) for game $\mathcal{G}_2$, and using the interference level caused by

$$\mathbf{Q}_{-q}^{(\boldsymbol{\tau}^q(n))} \triangleq \left(\mathbf{Q}_1^{(\tau_1^q(n))}, \ldots, \mathbf{Q}_{q-1}^{(\tau_{q-1}^q(n))}, \mathbf{Q}_{q+1}^{(\tau_{q+1}^q(n))}, \ldots, \mathbf{Q}_Q^{(\tau_Q^q(n))}\right). \quad (25)$$

Some standard conditions in asynchronous convergence theory that are fulfilled in any practical implementation need to be satisfied by the schedule $\{\tau_r^q(n)\}$ and $\{\mathcal{T}_q\}$; we refer to [13] for the details. Using the above notation, the asynchronous MIMO IWFA is formally described in Algorithm 1 below, where the mapping in (27) is defined as

$$\mathbf{T}_q(\mathbf{Q}_{-q}) \triangleq \tilde{\mathbf{WF}}_q\left(\tilde{\mathbf{H}}_{qq}^H \mathbf{R}_{-q}^{-1} \tilde{\mathbf{H}}_{qq}\right), \quad q = 1, \cdots, Q, \quad (26)$$

with $\tilde{\mathbf{WF}}_q(\cdot)$ given in (12) if the algorithm is applied to game $\mathcal{G}_1$, and it is defined as

$$\mathbf{T}_q(\mathbf{Q}_{-q}) \triangleq \mathbf{G}_q^{\sharp H} \overline{\mathbf{WF}}_q\left(\overline{\mathbf{H}}_{qq}^H \mathbf{R}_{-q}^{-1} \overline{\mathbf{H}}_{qq}\right) \mathbf{G}_q^{\sharp}, \quad q = 1, \cdots, Q,$$

with $\overline{\mathbf{WF}}_q(\cdot)$ given in (23) if the algorithm is applied to game $\mathcal{G}_2$. The mapping $\mathbf{T}_q(\mathbf{Q}_{-q})$ reduces to the classical MIMO waterfilling solution [19] if games $\mathcal{G}_\alpha$ and $\mathcal{G}_\infty$ are considered.

**Algorithm 1: MIMO Asynchronous IWFA**

Set $n = 0$ and $\mathbf{Q}_q^{(0)} =$ `any feasible point`;

`for` $n = 0 : \mathrm{N_{it}}$

$$\mathbf{Q}_q^{(n+1)} = \begin{cases} \mathbf{T}_q\left(\mathbf{Q}_{-q}^{(\boldsymbol{\tau}^q(n))}\right), & \text{if } n \in \mathcal{T}_q, \\ \mathbf{Q}_q^{(n)}, & \text{otherwise;} \end{cases} \qquad \forall q = 1, \cdots, Q \quad (27)$$

`end`

Convergence of the asynchronous IWFA is studied in [13, 14] (see also [11, 12] for special cases of the algorithm), where it was proved that the algorithm converges to the NE of the proposed games under the same conditions guaranteeing the uniqueness of the equilibrium. The proposed asynchronous IWFA contains as special cases a plethora of algorithms, each one obtained by a possible choice of the schedule $\{\tau_r^q(n)\}$, $\{\mathcal{T}_q\}$. The sequential [2, 11, 27, 28] and simultaneous [11]-[13] IWFAs are just two examples of the proposed general framework. The important result stated in [11]-[13] is that all the algorithms resulting as special cases of the asynchronous MIMO IWFA are guaranteed to reach the unique NE of game under the same set of convergence conditions, since convergence conditions do not depend on the particular choice of $\{\mathcal{T}_q\}$ and $\{\tau_r^q(n)\}$ [13].

Moreover all the algorithms obtained from Algorithm 1 have the following desired properties:

- ***Low complexity and distributed nature***: Even in the presence of null and/or shaping constraints, the best response of each user $q$ can be efficiently and locally computed using a MIMO waterfilling based solution, provided that each channel $\mathbf{H}_{qq}$ is replaced by the modified channel $\tilde{\mathbf{H}}_{qq}$ (if game $\mathcal{G}_1$ is considered) or $\overline{\mathbf{H}}_{qq}$ (if game $\mathcal{G}_2$ is considered). Thus, Algorithm 1 can be implemented in a distributed way, since each user only needs to measure the overall interference-plus-noise covariance matrix $\mathbf{R}_{-q}$ and waterfill over $\tilde{\mathbf{H}}_{qq}^H\mathbf{R}_{-q}^{-1}\tilde{\mathbf{H}}_{qq}$ [or over $\overline{\mathbf{H}}_{qq}^H\mathbf{R}_{-q}^{-1}\overline{\mathbf{H}}_{qq}$].

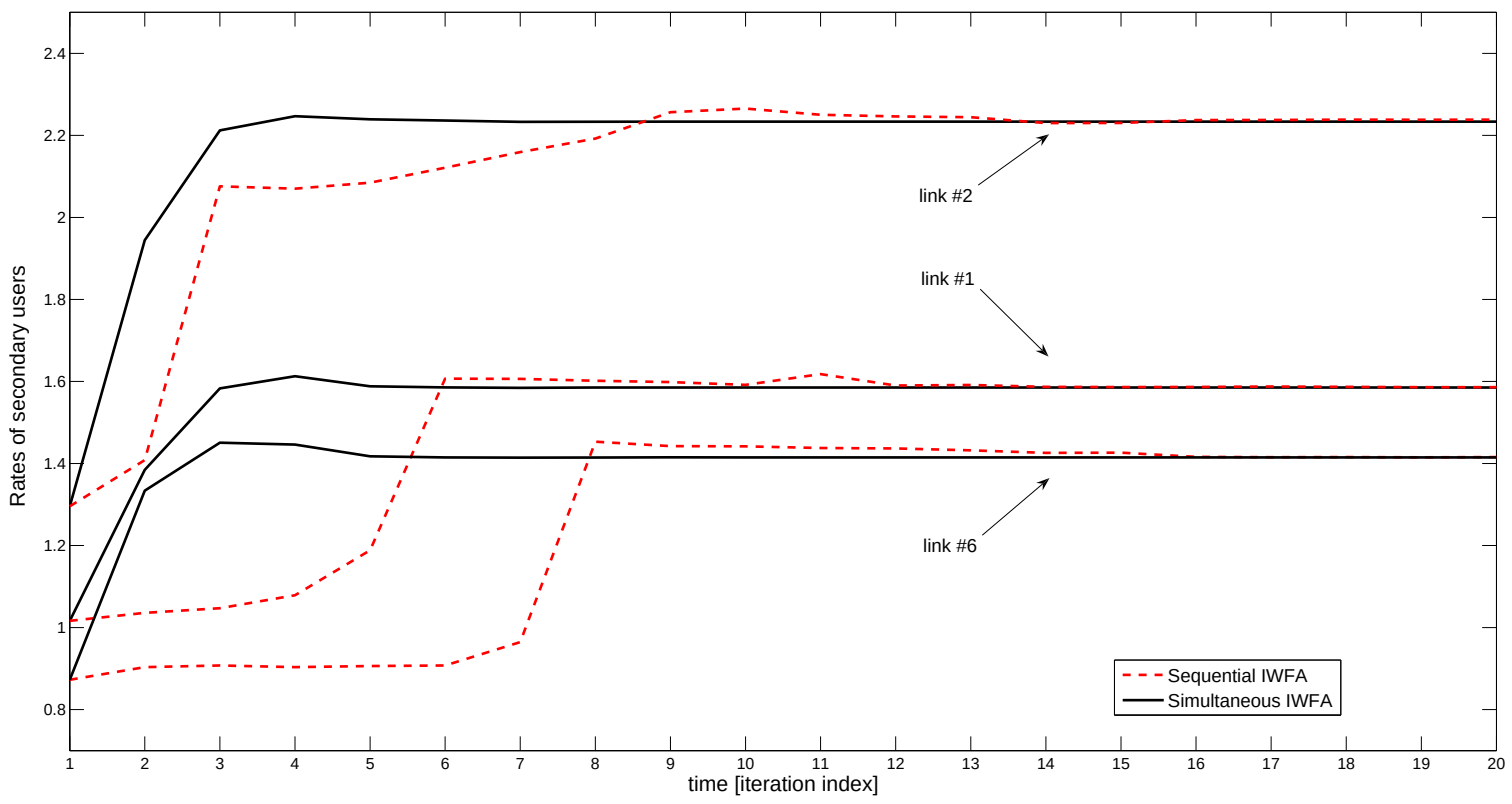

Figure 3: Simultaneous vs. sequential IWFA: rates of secondary users versus iterations, obtained by the sequential IWFA (dashed-line curves) and simultaneous IWFA (solid-line curves).

- ***Robustness***: Algorithm 1 is robust against missing or outdated updates of secondary users. This feature strongly relaxes the constraints on the synchronization of the users' updates with respect to those imposed, for example, by the simultaneous or sequential updating schemes [11]-[13].

- ***Fast convergence behavior:*** The simultaneous version of the proposed algorithm converges in a very few iterations, even in networks with many active secondary users. As an example, in Figure 3 we show the rate evolution of the of 3 links out 8 secondary users, corresponding to the sequential IWFA and simultaneous IWFA as a function of the iteration index. As expected, the sequential IWFA is slower than the simultaneous IWFA, especially if the number of active secondary users is large, since each user is forced to wait for all the users scheduled in advance, before updating his own covariance matrix. This intuition is formalized in [11], where the authors provided the expression of the asymptotic convergent factor of both the sequential and simultaneous IWFAs.

- ***Control of the radiated interference:*** Thanks to the game theoretical formulation including null and/or soft shaping constraints, the proposed asynchronous IWFA does not suffer of the main drawback of the classical sequential IWFA [27], i.e., the violation of the interference temperature limits [2].

Figure 4 shows an example of the optimal resource allocation based on the game theoretical formulation $\mathcal{G}_1$, for a cognitive MIMO network composed by two primary users and two secondary users, sharing the same spectrum and space. Secondary users are equipped with four transmit/receive antennas, placed in uniform linear arrays critically spaced at half of the wavelength of the passband transmitted signal. For the sake of simplicity, we assumed that the channels between the transmitter and the receiver of the secondary users have three physical paths (one line-of-sight and two reflected paths) as shown in Figure 4(a). To preserve the QoS of primary users' transmissions, null constraints are imposed to secondary users in the (line-of-sight) "directions" of primary users' receivers [see subplot (a)]. For the scenario shown in the figure, one null constraint for each player is imposed along the transmit directions $\phi_1 = \pi/2$ and $\phi_2 = -5\pi/12$. This can be done choosing for each player $q$ the matrix $\mathbf{U}_q$ in (21) coinciding with the spatial signature vector in the transmit direction $\phi_q$, i.e., $\mathbf{U}_q = [1, \exp(-j2\pi\Delta_{t_q}\sin(\phi_q)), \exp(-j2\pi 2\Delta_{t_q}\sin(\phi_q)), \exp(-j2\pi 3\Delta_{t_q}\sin(\phi_q))]^T$, with $\Delta_{t_q} = 1/2$ denoting the normalized (by the signal wavelength) transmit antenna separation and $q = 1, 2$. In Figure 4(b), we plot the transmit beamforming patterns, associated to the eigenvectors of the optimal covariance matrix of the two secondary users at the NE, obtained using Algorithm 1. In each radiation diagram plot, solid (blue) and dashed (black) line curves refer to the two eigenvectors corresponding to the nonzero eigenvalues (arranged in increasing order) of the optimal covariance matrix [recall that, because of the null constraints, the equivalent channel matrix $\widetilde{\mathbf{H}}_{qq}$ in (21) has rank equal to 2]. Observe that the null constraints guarantee that at the NE no power is radiated by the two secondary transmitters along the directions $\phi_1$ (for transmitted one) and $\phi_2$ (for transmitted two), showing that in the MIMO case, the orthogonality among primary and secondary users can be reached in the space rather than in the frequency domain, implying that primary and secondary users may share frequency bands, if this is allowed by FCC spectrum policies.

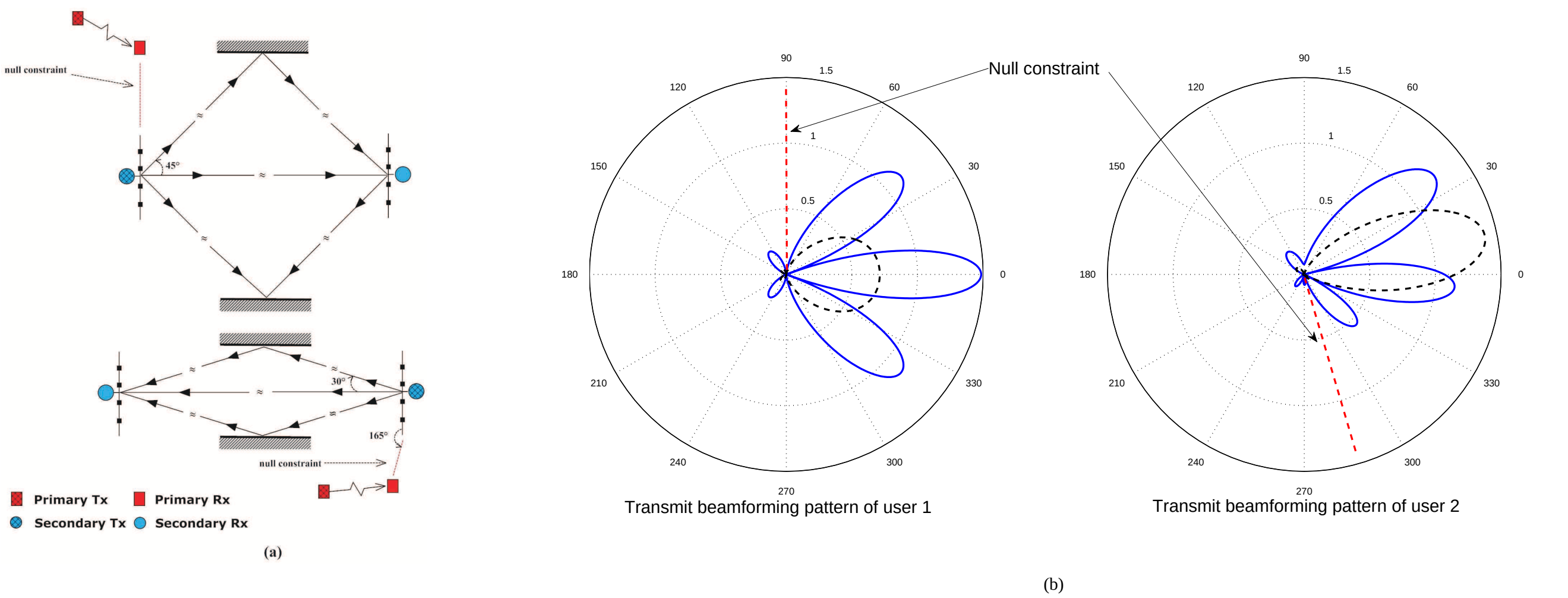

Figure 4: Optimal transmit beamforming patterns at the NE of game $\mathcal{G}_1$ [subplot (b)] for a cognitive MIMO network composed by two primary and two secondary users [subplot (a)].

# 5 Special Cases

The MIMO game theoretic formulation proposed in the previous sections provides a general and unified framework for studying the resource allocation problem based on rate maximization in hierarchical CR networks, where primary and secondary users coexist. In this section, we specialize the results to two scenarios of interest: 1) the spectrum sharing problem among primary and secondary users transmitting over *SISO frequency-selective channels*; and 2) the MIMO transceivers design of heterogeneous systems sharing the same spectrum over unlicensed bands.

## 5.1 Spectrum sharing over SISO frequency-selective channels with spectral mask constraints

The block transmission over SISO frequency-selective channels is obtained from the I/O model in (1), when each channel matrix $\mathbf{H}_{rq}$ is a $N \times N$ Toeplitz circulant matrix, $\mathbf{R}_{n_q}$ is a $N \times N$ diagonal matrix $N$ is the length of the transmitted block (see, e.g., [10]). This leads to the following eigendecomposition for each channel $\mathbf{H}_{rq} = \mathbf{W}\mathbf{D}_{rq}\mathbf{W}^H$, where $\mathbf{W}$ is the normalized IFFT matrix, i.e., $[\mathbf{W}]_{ij} \triangleq e^{j2\pi(i-1)(j-1)/N}/\sqrt{N}$ for $i, j = 1, \ldots, N$ and $\mathbf{D}_{rq}$ is a $N \times N$ diagonal matrix, where $[\mathbf{D}_{rq}]_{kk} \triangleq H_{rq}(k)$ is the frequency-response of the channel between source $r$ and destination $q$. Within this setup, we focus on game $\mathcal{G}_1$ given in (8), but similar results could be obtained if game $\mathcal{G}_2$, $\mathcal{G}_\alpha$ or $\mathcal{G}_\infty$ were considered instead. In the case of SISO frequency-selective channels, game $\mathcal{G}_1$ can be rewritten as:

$$\begin{array}{lll} \underset{\mathbf{Q}_q \succeq \mathbf{0}}{\text{maximize}} & \log\det\left(\mathbf{I} + \mathbf{H}_{qq}^H \mathbf{R}_{-q}^{-1} \mathbf{H}_{qq} \mathbf{Q}_q\right) & \\ \text{subject to} & \mathsf{Tr}(\mathbf{Q}_q) \leq P_q & \forall q = 1, \cdots, Q, \\ & \left[\mathbf{W}^H \mathbf{Q}_q \mathbf{W}\right]_{kk} \leq p_q^{\max}(k), \quad \forall k = 1, \cdots, N, & \end{array} \tag{28}$$

where $\{p_q^{\max}(k)\}$ is the set of spectral mask constraints, that can be used to impose shaping (and thus also null) constraints on the transmit power spectral density (PSD) of secondary users over licensed/unlicensed bands.

**Nash equilibria:** The solutions of the game in (28) have the following structure [10]:

$$\mathbf{Q}_q^\star = \mathbf{W}\,\mathrm{Diag}(\mathbf{p}_q^\star)\mathbf{W}^H, \quad \forall q = 1, \cdots, Q, \tag{29}$$

where $\mathbf{p}_q^\star \triangleq (p_q^\star(k))_{k=1}^N$ is the solution to the following set of fixed-point equations

$$\mathbf{p}_q^\star = \mathsf{wf}_q(\mathbf{p}_{-q}^\star)\ , \quad \forall q = 1, \cdots, Q, \tag{30}$$

with the waterfilling vector operator $\mathsf{wf}_q(\cdot)$ defined as

$$[\mathsf{wf}_q(\mathbf{p}_{-q})]_k \triangleq \left[\mu_q - \frac{1 + \sum_{r \neq q} |H_{rq}(k)|^2 p_r(k)}{|H_{qq}(k)|^2}\right]_0^{p_q^{\max}(k)}, \quad k = 1, \cdots, N, \tag{31}$$

where $\mu_q$ is chosen to satisfy the power constraint with equality $\sum_k p_q^\star(k) = P_q$.

Equation (29) states that, in the case of SISO frequency-selective channels, a NE is reached using, for each user, a multicarrier strategy (i.e., the diagonal transmission strategy through the frequency bins), with a proper power allocation. This simplification with respect to the general MIMO case, is a consequence of the property that all channel Toeplitz circulant matrices are diagonalized by the *same* matrix, i.e., the IFFT matrix $\mathbf{W}$, that does not depend on the channel realization.

Interestingly, multicarrier transmission with a proper power allocation for each user is still the optimal transmission strategy if in (28) instead of the information rate, one considers the maximization of the transmission rate using finite order constellations and under the same constraints as in (28) plus a constraint on the average error probability. Using the gap approximation analysis, the optimal power allocation is still given by the waterfilling solution (31), where each channel transfer function $|H_{qq}(k)|^2$ is replaced by $|H_{qq}(k)|^2/\Gamma_q$, where $\Gamma_q \geq 1$ is the gap [10]. The gap depends only on the family of constellation and on error probability constraint $P_{e,q}$; for $M$-QAM constellations, for example, the resulting gap is $\Gamma_q = (\mathcal{Q}^{-1}(P_{e,q}/4))^2/3$ (see, e.g., [29]).

Reaching a NE of the game in (28) satisfies a competitive optimality principle, but, in general, multiple equilibria may exist, so that one is never sure about which equilibrium is really reached. Sufficient conditions on the MUI that guarantee the uniqueness of the equilibrium have been proposed in the literature [10]-[12] and [27, 28]. Among all, one of the two following conditions is sufficient for the uniqueness of the NE:

$$\sum_{r\neq q} \max_k \frac{\left|\bar{H}_{rq}(k)\right|^2}{\left|\bar{H}_{qq}(k)\right|^2}\frac{d_{qq}^2}{d_{rq}^2} < 1, \ \forall q = 1,\cdots,Q, \quad \text{and} \quad \forall k = 1,\cdots,N, \tag{32}$$

$$\sum_{r\neq q} \max_k \frac{\left|\bar{H}_{rq}(k)\right|^2}{\left|\bar{H}_{qq}(k)\right|^2}\frac{d_{qq}^2}{d_{rq}^2} < 1, \quad \forall r = 1,\cdots,Q, \quad \text{and} \quad \forall k = 1,\cdots,N, \tag{33}$$

where we have introduced the normalized channel transfer functions $H_{rq}(k) \triangleq \bar{H}_{rq}(k)/d_{rq}^2$, $\forall r, q$, with $d_{rq}$ indicating the distance between transmitter of the $r$-th link and the receiver of the $q$-th link. From (32)-(33), it follows that, as expected, the uniqueness of NE is ensured if secondary users are sufficiently far apart from each other. In fact, from (32)-(33) for example, one infers that there exists a minimum distance beyond which the uniqueness of NE is guaranteed, corresponding to the maximum level of interference that may be tolerated by the users. Specifically, condition (32) imposes a constraint on the maximum amount of interference that each receiver can tolerate; whereas (33) introduces an upper bound on the maximum level of interference that each transmitter is allowed to generate. Interestingly, the uniqueness of the equilibrium does not depend on the interference generated by the transmissions of primary users.

**Asynchronous IWFA:** To reach the equilibrium of the game, secondary users can perform the asynchronous IWFA based on the mapping in (31). This algorithm can be obtained directly from Algorithm 1, as special case. It was proved in [12] that, e.g., under conditions (32)-(33), the asynchronous IWFA based on mapping (31) converges to the unique NE of game in (28) as Nit$\rightarrow \infty$, for any set of feasible initial conditions and updating schedule.

In Figure 5, we show an example of the optimal power allocation in SISO frequency-selective channels at the NE, obtained using the proposed asynchronous IWFA, for a CR system composed by one primary user [subplot (a)] and two secondary users [subplot (b)], subject to null constraints over licensed bands, spectral mask constraints and transmit power constraints. In each plot, solid and dashed-dot line curves refer to optimal PSD of each link and PSD of the MUI plus thermal noise, normalized by the channel transfer function square modulus of the link, respectively. In this example, there is a band $A$ (from 50 to 300 frequency bins) allocated to an active primary user; there is then a band $B$ (from 300 to 400 frequency bins) allocated to licensed users, but temporarily unused; the rest of the spectrum, denoted as $C$, is vacant. The temporarily void band $B$ can be utilized by secondary users, provided that they do not overcome a maximum tolerable spectral density. The optimal power allocations shown in Figure 5 are the result of running the simultaneous IWFA. We can observe that the secondary users do not transmit over band $A$ and they allocate their power over both bands $B$ and $C$, respecting a power spectral density limitation over band $B$.

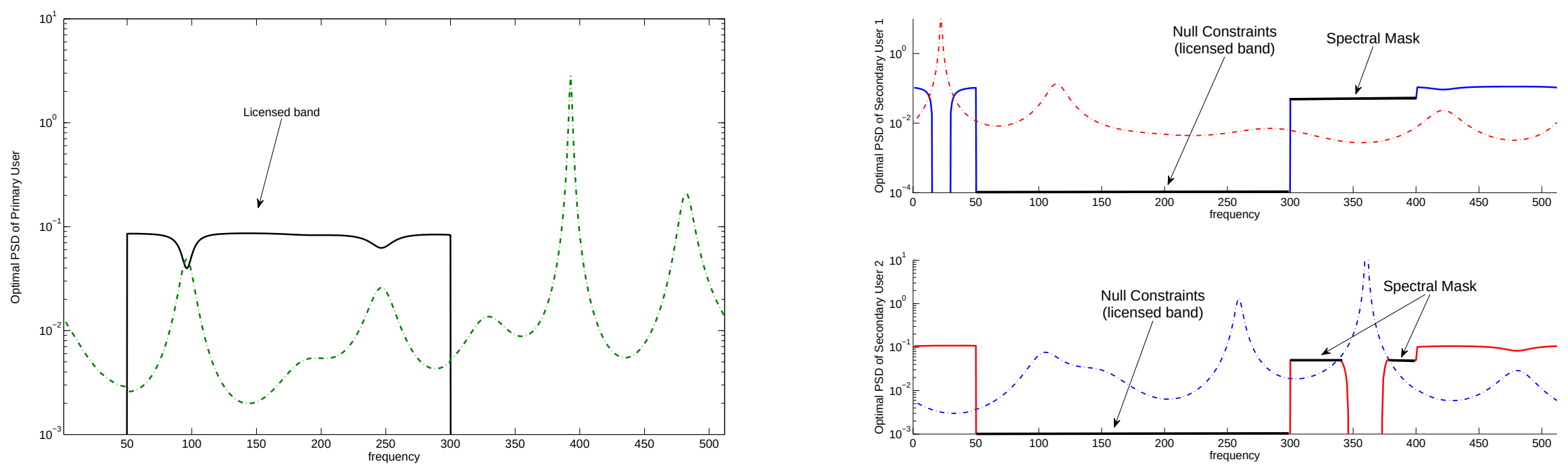

(a) (b)

Figure 5: Spectrum sharing among one primary [subplot (a)] and two secondary users [subplot (b)]: Optimal PSD of each link (solid lines), and PSD of the MUI-plus-thermal noise normalized by the channel transfer function square modulus of the link (dashed-dot line).

## 5.2 MIMO transceiver design of heterogeneous systems in unlicensed bands

We consider now on a scenario where multiple unlicensed MIMO cognitive users share the same unlicensed spectrum and geographical area. The availability of MIMO transceivers clearly enriches the possibilities for spectrum sharing as it adds the extra spatial degrees of freedom. In unlicensed bands, there are no interference constraints to be satisfied by the users. Thus, the game theoretical formulation as given in (8), without considering the null constraints, seems the most appropriate to study the resource allocation problem in this scenario. In the following we refer to game $\mathcal{G}_1$ assuming tacitly that the null constraints are removed.

Similarly to the SISO case, sufficient conditions for the uniqueness of the NE are given by one of the two following set of conditions (more general conditions are given in [13]):

$$\text{Low MUI received:} \qquad \sum_{r\neq q} \rho\left(\mathbf{H}_{rq}^{H}\mathbf{H}_{qq}^{-H}\mathbf{H}_{qq}^{-1}\mathbf{H}_{rq}\right) < 1, \quad \forall q = 1, \cdots, Q, \tag{34}$$

$$\text{Low MUI generated:} \qquad \sum_{q\neq r} \rho\left(\mathbf{H}_{rq}^{H}\mathbf{H}_{qq}^{-H}\mathbf{H}_{qq}^{-1}\mathbf{H}_{rq}\right) < 1, \quad \forall r = 1, \cdots, Q. \tag{35}$$

Conditions (34)-(35) quantify how much MUI can be tolerated by the systems to guarantee the uniqueness of the NE. Interestingly, (32)-(33) and most of the conditions known in the literature [11, 27, 28] for the uniqueness of the NE of the rate-maximization game in SISO frequency-selective interference channels and OFDM transmission come naturally from (34)-(35) as special cases.

The Nash equilibria of game $\mathcal{G}_1$ can be reached using the asynchronous IWFA described in Algorithm 1, whose convergence is guaranteed under conditions (34)-(35), for any set of initial conditions and updating schedule of the users.

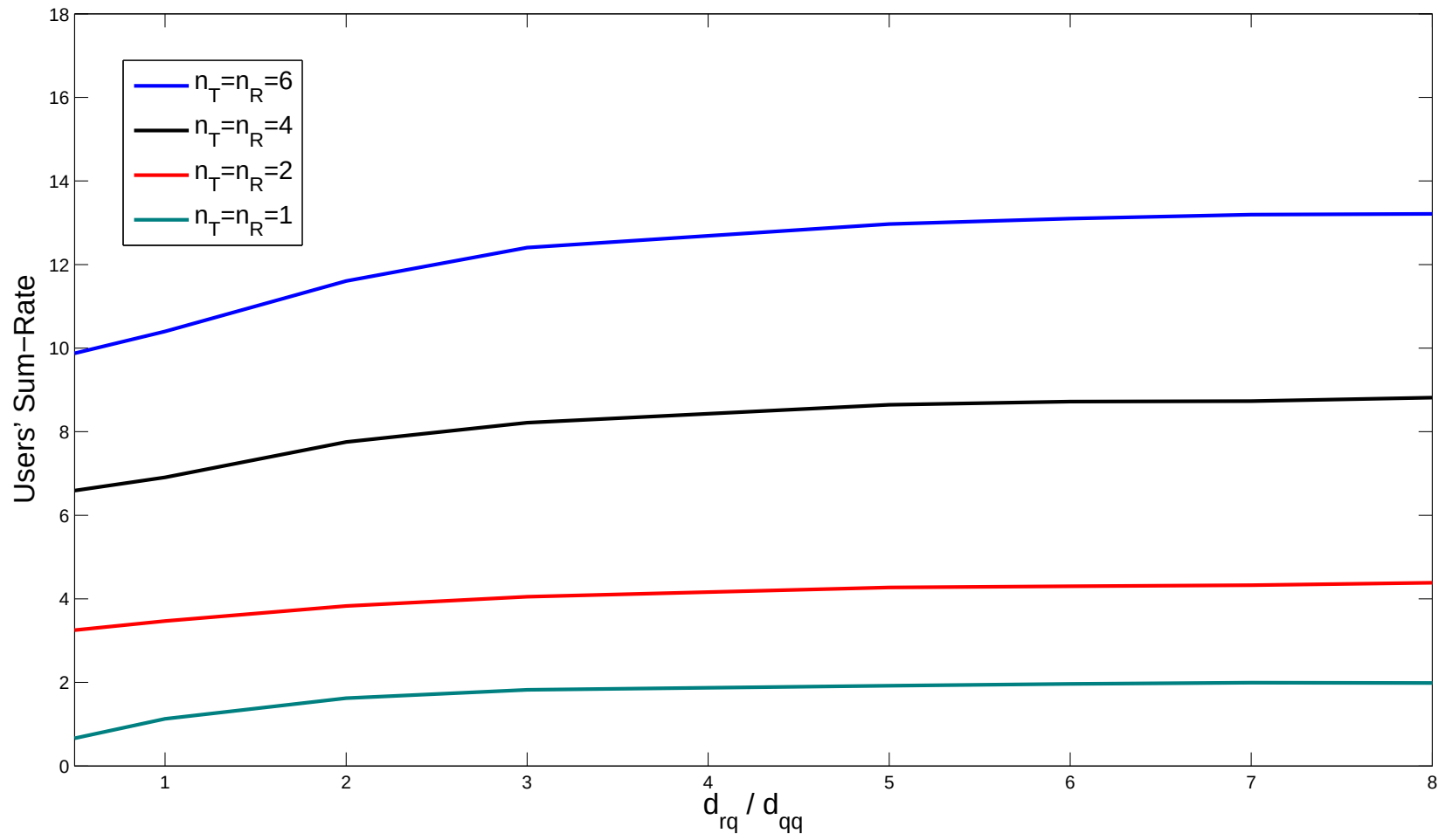

Figure 6: Sum-Rate of the users versus the inter-pair distance $d_{rq}/d_{qq}$; $d_{rq} = d_{qr}$, $d_{rr} = d_{qq} = 1$, $\forall r, q$, for different numbers of transmit/receive antennas.

In Figure 6 we show an example of the benefits of MIMO transceivers in the cognitive radio context. We plot in the figure the sum-rate of a two-user frequency-selective MIMO system as a function of the inter-pair distance among the links, for different number of transmit/receive antennas. The rate curves are averaged over 500 independent channel realizations, whose taps are simulated as i.i.d. Gaussian random variables with zero mean and unit variance. For the sake of simplicity, the system is assumed to be symmetric, i.e., the transmitters have the same power budget and the interference links are at the same distance (i.e., $d_{rq} = d_{qr},\ \forall q, r$), so that the cross channel gains are comparable in average sense. From the figure one infers that, as for isolated single-user systems or multiple access/broadcast channels, also in MIMO interference channels, increasing the number of antennas at both the transmitter and the receiver side leads to a better performance. The interesting result, coming from Figure 6, is that the incremental gain due to the use of multiple transmit/receive antennas is almost independent of the interference level in the system, since the MIMO (incremental) gains in the high-interference case (small values of $d_{rq}/d_{qq}$) almost coincide with the corresponding (incremental) gains obtained in the low-interference case (large values of $d_{rq}/d_{qq}$), at least for the system simulated in Figure 6. This desired property is due to the fact that the MIMO channel provides more degrees of freedom for each user than those available in the SISO channel, that can be explored to find out the best partition of the available resources for each user, possibly cancelling the MUI.

## 6 Conclusion and Directions for Further Developments

In this paper we have proposed a signal processing approach to the design of CR systems, using a competitive optimality principle, based on game theory. We have addressed and solved some of the challenging issues in CR, namely: 1) the establishment of conditions guaranteeing that the dynamical interaction among cognitive nodes, under constraints on the transmit spectral mask and on interference induced to primary users, admits a (possibly unique) equilibrium; and 2) the design of decentralized algorithms able to reach the equilibrium points, with minimal coordination among the nodes. We have seen how basic signal processing tools such as subspace projectors play a fundamental role. The spectral mask constraints have been in fact used in a very broad sense, meaning that the projection of the transmitted signal along prescribed subspaces should be null (null constraints) or below a given threshold (soft constraints). The conventional spectral mask constraints can be seen as a simple case of this general set-up, valid for SISO channels and using as subspaces the space spanned by the IFFT vectors with frequencies falling in the guard bands. This general setup encompasses multiantenna MIMO systems, which is particularly useful for CR, as it provides the additional spatial degrees of freedom to control the interference generated by the cognitive users.

Of course, this field of research is full of interesting further directions worth of investigation. The NE points derived in this paper were dictated by the need of finding totally decentralized algorithms with minimal coordination among the nodes. However, the NE points may not be Pareto-efficient. This raises the issue of how to move from the NE towards the Pareto optimal trade-off surface, still using a decentralized approach. Game theory itself provides a series of strategies to move from inefficient Nash equilibria towards Pareto-efficient solutions, still using a decentralized approach, through, for example, repeated games, where the players learn from their past choices [9]. Examples of such games are the *auction* games, where the auctioneer (primary users) dynamically determine resource allocation and prices for the bidders (secondary users), depending on traffic demands, QoS and supply/demand curves, as evidenced in a series of works (see, e.g., [30, 31, 32]). Repeated games may also take the form of negotiations between primary and secondary users, with primary users willing to lease part of their spectrum to secondary users, under suitable remunerations [16] or under the availability given by secondary users to establish cooperative links with the primary users to improve their QoS [33]. Competitive pricing for spectrum sharing was also proposed as an oligopoly market where a few primary users offer spectrum access opportunities to secondary users [34]. An interesting issue will be the integration of our asynchronous IWFA in repeated (auction) games, where the optimization considers a set of primary users offering the lease of portion of their resources to a set of secondary users, as a function of traffic demands, QoS requirements and physical constraints.

Our search for the uniqueness conditions of the NE and the convergence conditions of our proposed algorithms forced us to simplify the model. For example, we assumed that each receiver has an error-free short-term prediction of the channel. This assumption was necessary for the mathematical tractability of the problem and to be able to provide closed-form expressions of our findings. This is useful to gain a full understanding of the problem, without relying on simulation results only. However, in practice, the transmitter is only able to acquire an estimate affected by errors and, based on that, to form a prediction of the short term future evolution. An interesting extension of the presented approach consists then in taking into account the effects of estimation errors and developing robust strategies. This is particularly relevant in CR systems because the strategy adopted by the cognitive users may be more or less aggressive depending on the reliability of their channel sensing.

Channel identification has a long history in signal processing. The problem becomes especially challenging in CR networks, where the estimation of the channel voids, for example, must be very accurate. Nevertheless, the estimation itself may be improved by exploiting the availability of a network of nodes that could, in principle, cooperate to get better and better estimates of the electromagnetic environment, working as a sensor network of cognitive nodes.

# References

[1] FCC Spectrum Policy Task Force, "FCC Report of the Spectrum Efficiency Working Group," Nov. 2002. `http://www.fcc.gov/sptf/files/SEWGFinalReport1.pdf`.

[2] S. Haykin, "Cognitive Radio: Brain-Empowered Wireless Communications,"*IEEE Jour. on Selected Areas in Communications*, vol. 23, no. 2, pp. 201-220, February 2005.

[3] I. F. Akyildiz, W.-Y. Lee, M. C. Vuran, S. Mohanty, "NeXt Generation/Dynamic Spectrum Access/Cognitive Radio Wireless Networks: A Survey," *Computer Networks*, vol. 50, pp. 2127–2159, 2006.

[4] Q. Zhao and B. Sadler, "A Survey of Dynamic Spectrum Access," *IEEE Signal Processing Magazine,*vol. 24, no. 3, pp. 79–89, May 2007.

[5] A. Goldsmith, S. A. Jafar, I. Maric, S. Srinivasa, "Breaking Spectrum Gridlock with Cognitive Radios: An Information Theoretic Perspective," to appear on *IEEE Jour. of Selected Areas in Communications (JSAC)*.

[6] J. Mitola, "Cognitive Radio for Flexible Mobile Multimedia Communication," in Proc. of the *IEEE International Workshop on Mobile Multimedia Communications (MoMuC) 1999*, pp. 3–10, November 1999.

[7] N. Devroye, P. Mitran, V. Tarokh, "Achievable Rates in Cognitive Radio Channels," *IEEE Trans. on Information Theory*, vol. 52, pp. 1813–1827, May 2006.

[8] M. J. Osborne and A. Rubinstein, *A Course in Game Theory*, MIT Press, 1994.

[9] T. Basar and G. J. Olsder, *Dynamic Noncooperative Game Theory*, SIAM Series in Classics in Applied Mathematics, Philadelphia, January 1999.

[10] G. Scutari, D. P. Palomar, and S. Barbarossa, "Optimal Linear Precoding Strategies for Wideband Non-Cooperative Systems based on Game Theory-Part I: Nash Equilibria," *IEEE Trans. on Signal Processing,* vol. 56, no. 3, pp. 1230–1249, March 2008.

[11] G. Scutari, D. P. Palomar, and S. Barbarossa, "Optimal Linear Precoding Strategies for Wideband Non-Cooperative Systems based on Game Theory-Part II: Algorithms," *IEEE Trans. on Signal Processing,* vol. 56, no. 3, pp. 1250–1267, March 2008.

[12] G. Scutari, D. P. Palomar, and S. Barbarossa, "Asynchronous Iterative Waterfilling for Gaussian Frequency-Selective Interference Channels," in *IEEE Trans. on Information Theory*, vol. 54, no. 7, pp. 2868–2878, July 2008.

[13] G. Scutari, D. P. Palomar, and S. Barbarossa, "Competitive Design of Multiuser MIMO Systems Based on Game Theory: A Unified View," *IEEE Jour. of Selected Areas in Communications (JSAC)*, special issue on "Game Theory in Communication Systems", vol. 26, no.7, September 2008.

[14] G. Scutari, D. P. Palomar, and S. Barbarossa, "MIMO Cognitive Radio: A Game Theoretical Approach,"submitted to IEEE Trans. on Signal Processing (2008). See also Proc. of the *9th IEEE Workshop on Signal Processing Advances for Wireless Communications* (*SPAWC 08*), Recife, Pernambuco, Brazil, July 6-9, 2008.

[15] Q. Zhao, "Spectrum Opportunity and Interference Constraint in Opportunistic Spectrum Access," in Proc. of *IEEE International Conference on Acoustics, Speech, and Signal Processing (ICASSP),* Honolulu, Hawaii, USA, April 15-20, 2007.

[16] L. Cao, H. Zheng, "Distributed Spectrum Allocation via Local Bargaining," in Proc. of *IEEE SECON 2005*, pp. 475–486, Sept. 2005.

[17] Y. Chen, Q. Zhao, A. Swami, "Joint design and separation principle for opportunistic spectrum access in the presence of sensing errors,"*IEEE Trans. on Information Theory*, vol. 54, pp. 2053 - 2071, May 2008.

[18] B. Wild and K. Ramchandran, "Detecting primary receivers for cognitive radio applications,"in Proc. of the *IEEE Symp. New Frontiers Dynamic Spectrum Access Networks* (*DYSPAN 05*), pp. 124–130, Nov. 2005.

[19] T. M. Cover and J. A. Thomas, *Elements of Information Theory*, John Wiley and Sons, 1991.

[20] W. Yu and R. Lui, "Dual methods for nonconvex spectrum optimization of multicarrier systems," *IEEE Trans. Commun.*, vol. 54, pp. 1310–1322, 2006.

[21] R. Cendrillon, W. Yu, M. Moonen, J. Verliden, and T. Bostoen, "Optimal multi-user spectrum management for digital subscriber lines," *IEEE Trans. Commun.,* vol. 54, no. 5, pp. 922–933, May 2006.

[22] S. Ye and R. S. Blum, "Optimized Signaling for MIMO Interference Systems with Feedback,"*IEEE Trans. on Signal Processing*, vol. 51, no. 11, pp. 2839-2848, November 2003.

[23] Y. Rong and Y. Hua, "Optimal power schedule for distributed MIMO links," in Proc. of the *25th Army Science Conference,* November 2006.

[24] Z.-Q. Luo and S. Zhang, "Dynamic Spectrum Management: Complexity and Duality," *IEEE Jour. of Selected Topics in Signal Processing*, vol. 2, no. 1, pp. 57–72, February 2008.

[25] R. A. Horn and C. R. Johnson, *Matrix Analysis*, Cambridge Univ. Press, 1985.

[26] Daniel P. Palomar and J. Fonollosa, "Practical Algorithms for a Family of Waterfilling Solutions," *IEEE Trans. on Signal Processing*, vol. 53, no. 2, pp. 686–695, Feb. 2005.

[27] W. Yu, G. Ginis, and J. M. Cioffi, "Distributed Multiuser Power Control for Digital Subscriber Lines," *IEEE JSAC*, vol. 20, no. 5, pp. 1105-1115, June 2002.

[28] Z.-Q. Luo and J.-S. Pang, "Analysis of Iterative Waterfilling Algorithm for Multiuser Power Control in Digital Subscriber Lines," *EURASIP Jour. on Applied Signal Processing*, May 2006.

[29] J. G. D. Forney and M. V. Eyuboglu, "Combined Equalization and Coding Using Precoding,"*IEEE Communications Magazine*, vol. 29, no. 12, pp. 25–34, December 1991.

[30] Y.-C. Liang, H.-H. Chen, J. Mitola, P. Mahonen, R. Kohno, J. H. Reed Eds., "Special Issue on Cognitive Radio: Theory and Application," *IEEE Jour. of Selected Areas in Communications (JSAC)*, vol. 26, no. 1, Jan. 2008.

[31] A. Swami, R. A. Berry, A. M. Sayeed, V. Tarokh, Q. Zhao Eds., "Special Issue on Signal Processing and Networking for Dynamic Spectrum Access," *IEEE Jour. of Selected Topics in Signal Processing*, vol. 2, no. 1, Feb. 2008.

[32] Z. Ji, K. Ray Liu, "Dynamic Spectrum Sharing: A Game Theoretical Overview," *IEEE Communications Magazine*, vol. 45, no. 5, pp. 88–94, May 2007.

[33] O. Simeone, I. Stanojev, S. Savazzi, Y. Bar-Ness, U. Spagnolini, R. Pickholtz, "Spectrum leasing to cooperating ad hoc secondary networks," *IEEE Jour. of Selected Areas in Communications (JSAC)*, vol. 26, no. 1, pp. 203–213, Jan. 2008.

[34] D. Niyato, E. Hossain, "Competitive Pricing for Spectrum Sharing in Cognitive Radio Networks: Dynamic Game, Inefficiency of Nash Equilibrium, and Collusion," *IEEE Jour. of Selected Areas in Communications (JSAC)*, vol. 26, no. 1, pp. 192–202, Jan. 2008.